\documentclass[preprint,floats,aps,showpacs,epsf]{revtex4}
\usepackage{epsfig}

\def\be{\begin{equation}}
\def\ee{\end{equation}}
\def\bdm{\begin{displaymath}}
\def\edm{\end{displaymath}}
\def\bea{\begin{eqnarray}}
\def\eea{\end{eqnarray}}

\def\s{\sigma}

\newcommand{\p}{\partial}

\newcommand{\la}{\langle}
\newcommand{\ra}{\rangle}
\newcommand{\rd}{\mbox{d}}
\newcommand{\ri}{\mbox{i}}
\newcommand{\re}{\mbox{e}}

\begin{document}
\draft
\title{Confinement and deconfinement of spinons in a frustrated spin-1/2 Heisenberg  model. }
\author{A. M. Tsvelik }
\affiliation{ Department of  Physics, Brookhaven 
National Laboratory, Upton, NY 11973-5000, USA}
\date{\today}
\begin{abstract}
In this publication I discuss the phase diagram of a frustrated spin-1/2 Heisenberg model suggested in A. A. Nersesyan and A. M. Tsvelik, Phys. Rev. B{\bf 67}, 024422 (2003). The phase diagram contains $(\pi,0)$ and $(\pi,\pi)$ antiferromagnetic phases separated by the Valence Bond Crystal (VBC) state. I argue that the point of the  phase  diagram with deconfined  spinons, predicted in the aforementioned work,  is situated in the middle of VBC state, at the point where the dimerization order parameter changes sign. 
\end{abstract}
\maketitle

\section{Introduction}

 In our previous paper we described a model of spin-1/2 antiferromagnet (the so-called Confederate Flag or CF model) where the fine tuning of interactions gives  rise to a state with fractional quantum spin number excitations (spinons) \cite{nerstsv}. The subsequent studies (\cite{moukouri},\cite{sindz}, \cite{balst} and especially \cite{batista}) have provided a support to our results. In the original publication we did not discuss   what happens if one deviates from this special point. Such discussion is a subject of the present    
 publication. Here  I discuss the phase diagram of CF model concentrating primarily on the vicinity of the deconfinement (D) point. This gives us a better understaning of the physics involved and also helps to put the CF model in the broader context of studies  of frustrated magnetism. Though the corresponding literature is enormous (see, for example,  \cite{lhuillier} for review), the theoretical efforts are primarily concentrated on idealized models (such as models of dimers or gauge field theory models) whose relation to  microscopic models with realistic interactions is not clear. The standard argument invokes universality: the belief is  that low energy behavior of such systems will be independent on microscopical details following  some universal patterns. It is always interesting to check general considerations against concrete models. In that sense, CF model, being simply a model of a Heisenberg magnet with short range interactions,  presents an almost unique example. 

 The literature knows two scenario for realization of the  D-point. In one of them deconfined spinons exist  on the boundary between antiferromagnetic and VBC state in the hypothetical situation when these states touch each other at a Quantum Critical Point \cite{subir}. In the other scenario, based on the study of models of quantum dimers,  spinons  appear at the boundary between two VBC states (the Roksar-Kivelson critical point) \cite{fradkin}, \cite{sent}. As I shall argue  in this paper, neither situation is realized in CF model. In that model the D-point separates two  VBC states, as in the second scenario, but the analogy does not go much further. First,  according to \cite{fradkin},\cite{sent}, the spectrum at the D-point in the dimer models consists of  spinless particles (``photons'') with a quadratic $\omega \sim k^2$ spectrum. This is absolutely incompatible with CF model which is approximately (1+1) Lorentz invariant when the interchain coupling is weak. This symmetry dictates that the dispersion of the gapless excitations along the chains direction must be linear. Second, one of the VBC phases in the dimer models contains a 'devil's staircase' of commensurate and incommensurate phases which does not agree with the situation in  CF model where both VBC phases are simple. So it appears that CF model is quite distinct and does not fit into the known categories.

\section{The model}

The model suggested in our original paper \cite{nerstsv} is a spin-1/2
Heisenberg magnet with spatially  anisotropic exchange interactions. The exchange in one direction is much stronger than in the others and therefore this model can be viewed as a model of weakly coupled chains. Recently Batista and Trugman \cite{batista} have found an isotropic version of the CF  model and shown that it has the same ground state degeneracy and possesses spin-1/2 excitations. Therefore the requirement of the space anisotropy is only a matter of convenience. It allows us to use the continuum limit in one direction and to apply the field theory methods. The existence of the deconfined point also does not depend on the number of transverse directions, therefore for the sake of simplicity I will discuss the two-dimensional version of the model. In that case the interaction pattern reminds the Confederate Flag (see Fig. 1). 
 The CF model Hamiltonian is given by 
\bea
H_{CF} = \sum_{j,n}\left\{J_{\parallel}{\bf S}_{j,n} \cdot {\bf S}_{j+1,n} + 
\sum_{\mu = \pm 1} \left[ J_r {\bf S}_{j,n} + J_d
\left( {\bf S}_{j+1,n} + {\bf S}_{j-1,n} \right) \right] \cdot {\bf S}_{j,n + \mu} \right\} + \lambda V_{quarter},\label{CF}
\eea
 where ${\bf S}_{j,n}$ are spin-1/2 operators, and 
$J_{\parallel}>>  J_r , J_d > 0$ . The term $V_{quarter}$ contains a four-(and possibly higher) spin  exchange interaction with a small coupling constant $\lambda \sim J_r^2/J_{\parallel}$.  This term was absent in the original publication, but, as was recently demonstrated by Balents and Starykh \cite{balst}, one  needs to introduce it  
to fine-tune the model to the state with deconfined spinons.

\begin{figure}[ht]
\begin{center}
\epsfxsize=0.5\textwidth
\epsfbox{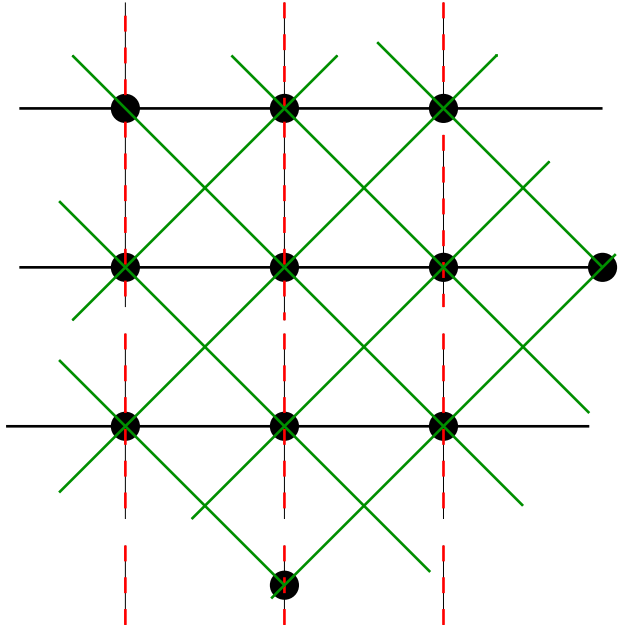}
\end{center}
\caption{ Exchange interactions pattern for CF model, the red lines correspond to $J_r$ and the green ones to $J_d$.}
\end{figure}

 Assuming that  the interchain couplings ($J_r, J_d, \lambda$) 
are  much smaller than the exchange along the chains ($J_{\parallel}$)
 it is legitimate to adopt  
a continuum description of individual chains. In this description,
the local spin densities are represented as sums of the smooth and  staggered parts:
\be
{\bf S}_{j,n} /a_0 \to   {\bf S}_n (x) = {\bf M}_n(x) + (-1)^j{\bf N}_n(x),
~~~x = j a_0,
\label{continS}
\ee 
$a_0$ being the lattice spacing in the chain direction.
 
\begin{figure}[ht]
\begin{center}
\epsfxsize=0.65\textwidth
\epsfbox{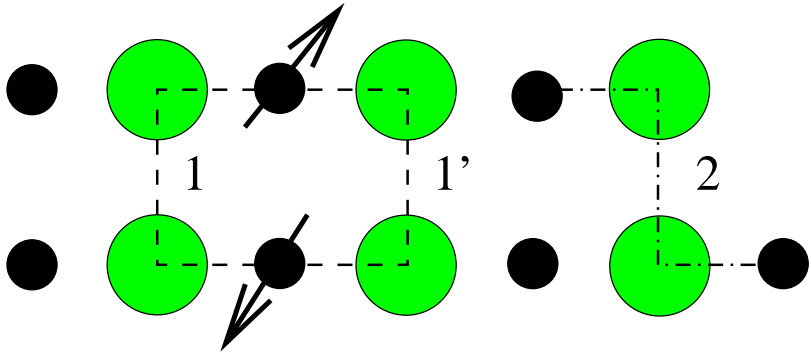}
\end{center}
\caption{ A possible realization of the Confederate Flag exchange pattern via a superexchange.  
The superexchange between  spins of magnetic (black) ions occurs through orbitals of non-magnetic (green) ones with  a big ionic radius. The pathes 1 and 1' generate the vertical interchain exchange ($J_r$) and the pathes of type 2 generate the diagonal one such that $J_g = J_r/2$. }
\end{figure}

The low-energy dynamics of the spin--1/2 Heisenberg antiferromagnet 
\bea
H_{1D} = J_{\parallel}\sum_j({\bf S}_j{\bf S}_{j + 1})
\eea
is 
described by the SU$_1$(2) Wess-Zumino-Novikov-Witten model.  
The latter  Hamiltonian can be written in terms of the so-called 
chiral vector {\it current} operators, 
${\bf J}$ and $\bar{\bf J}$, satisfying the level $k =1$ Kac-Moody algebra 
(this approach has been described in a 
vast number of publications; (for a review see \cite{affleck1} or \cite{Book},\cite{me}):
\bea
H_{1D} \to  \frac{2\pi v}{3}\int \rd x\left[:({\bf J}\cdot {\bf J}): + 
:({\bf \bar J}\cdot {\bf \bar J}):\right] + \cdots,
\label{WZNW}
\eea
with $v = \pi a_0J_{\parallel}/2$. 

It is remarkable that the smooth part of magnetization,
\be
{\bf M} = {\bf J} + \bar{\bf J}, \label{M}
\ee 
and the spin current,
\be
{\bf j} = v({\bf J} - \bar{\bf J}), \label{j}
\ee
are locally expressed in terms of the chiral currents.

In the CF model,
the exchange 
is frustrated in the direction perpendicular to the chains. In order to eliminate the coupling of the leading relevant operators (the  staggered magnetizations  
${\bf N}_n (x) \cdot {\bf N}_{n+1} (x)$ and the dimerizations $\epsilon_n(x)\epsilon_{n + \mu}(x)$) one has to fine tune the couplings $J_r - 2J_d$ and $\lambda$. In the leading order in $J_r, J_d$ the fine tuning is achieved   when $J_r - 2J_d =0, \lambda =0$.

 To discuss the phase diagram of CF model one needs to  deviate from the fine-tuned point with deconfined spinons. For weak interchain interactions one still can employ the continuum  Hamiltonian obtained using the continuum description of individual chains following 
the asymptotic representation (\ref{continS}) of the spin operators: 
\bea
&& H = H_1 + H_2, \\
&& H_1 = \sum_{n =1}^{2N} \left[H_{1D,n} + \frac{\gamma}{2}\sum_{\mu=\pm 1}({\bf J} + 
\bar{\bf J})_n \cdot ({\bf J} + \bar{\bf J})_{n + \mu}\right].\label{model}\\
&& H_2 = \frac{1}{2}\delta J \sum_{n, \mu} {\bf N}_{n}{\bf N}_{n +\mu}\label{V}
\eea
where $H_{1D}$ is given by Eq.(\ref{WZNW}). For the lattice Hamiltonian presented on Fig. 1 we have   $\gamma = J_r + 2J_d$ and $\delta J = J_r -2J_d$. One very important property of model (\ref{model}) is its (1+1)-dimensional Lorentz invariance which survives at $\delta J \neq 0$. This dictates the form of the  excitation spectrum for all particles:
\bea
E^2 = v^2k^2 + \Delta^2(k_{\perp}) 
\eea
where $\Delta(k_{\perp})$ is a periodic function of the wave vector component perpendicular to the chains.  I emphasise that this result is valid for an arbitrary number of chains. As I have mentioned in Introduction, this property precludes the existence of $k^2$ gapless modes characteristic for dimer model critical points.  

 Model (\ref{model},\ref{V}) can be viewed as the  critical system (the bunch of non-interacting spin-1/2 chains) perturbed by relevant interactions. Each of these interactions generate their own energy scale according to their scaling dimension. The current-current interaction, being only marginally relevant, generates the scale 
\be
M \sim J_{\parallel}\exp[- \pi^2 J_{\parallel}/2(J_r + 2J_d)] \label{Mass}
\ee 
and the interaction of the staggered spin components generates a scale $\sim \delta J$. When $|\delta J| >> M$ the system orders antiferromagnetically (though in two dimensions only at $T =0$). The corresponding Neel wave vectors are $(\pi,0)$ for $\delta J < 0$ and $(\pi,\pi)$ for $\delta J > 0$. These things are absolutely obvious; it is less obvious however what happens in the opposite limit $M >> |\delta J|$.  For weak interchain interactions the best I can do in this case is to study the four chain model,  where non-perturbative calculations can be carried out explicitely. I believe that the  four chains are representative enough to give an  insight into what happens for an infinite system.  

\section{The four chain model as both solvable and  representative case.}

 So let us consider the case of four chains with periodic boundary conditions in the transverse direction. Let me  briefly recall the results for  $\delta J =0$ obtained in \cite{nerstsv},\cite{smirnov}. As was noticed that at $\delta J =0$ the model acquires an additional symmetry: the sectors with different parity decouple. This follows from the fact that the relevant interactions couple  only currents of different chirality belonging to different chains. Thus the  relevant current-current  interaction is  
\bea
({\bf J}_1 + {\bf J}_3)(\bar{\bf J}_2 + \bar{\bf J}_4) + ({\bf J}_2 + 
{\bf J}_4)(\bar{\bf J}_1 + \bar{\bf J}_3) 
\eea
and the Hamiltonian (\ref{model}) decouples into two parts:
\bea
H = H_+ + H_-
\eea
The $+$ parity sector contains $J_{1,3}$ and $\bar J_{2,4}$ currents 
and the $-$ parity sector contains $\bar J_{1,3}$ and $J_{2,4}$ currents. Each of the models represented by the Hamiltonians $H_{\pm}$ is integrable, which  can be demonstrated by rewriting them in terms of familiar integrable models. Let us recall how it was done in \cite{nerstsv}. A  sum of two $k =1$ SU(2) currents is the $k=2$ current; moreover, according to \cite{FatZam} the sum of two SU$_1$(2) WZNW models (the central charge 2) can be represented as the SU$_2$(2) WZNW model with central charge 3/2 and plus one massless Majorana fermion (a critical Ising model) with central charge 1/2.   Using  the results of \cite{FatZam} we   rewrite the entire Hamiltonian density (\ref{WZNW}) as follows (here only the  (+)-parity part is written):
\bea
&&{\cal H}_+ = {\cal H}_{massless} + {\cal H}_{massive}\nonumber\\
&&{\cal H}_{massless} =  -\frac{\ri v}{2}\chi_{0}\p_x\chi_{0} + 
\frac{\ri v}{2}\bar\chi_{0}\p_x\bar\chi_{0}\label{Ising}\\
&&{\cal H}_{massive} = \frac{\pi v}{2}(:{\bf I}\cdot {\bf I}: + 
:\bar{\bf I}\cdot \bar{\bf I}:) + \gamma{\bf I}\cdot \bar{\bf I} = 
\frac{\ri v}{2}(- \chi^a\p_x\chi^a + \bar\chi^a\p_x\bar\chi^a) - 
\frac{\gamma}{2}(\chi^a\bar\chi^a)^2 \label{O3} 
\eea
where $a = 1,2,3$ and 
\[
{\bf I} = {\bf J}_1 + {\bf J}_3, ~~ \bar{\bf I} = \bar{\bf J}_2 + \bar{\bf J}_4
\]
\bea
{\bf J}_{1,3} = \frac{\ri}{2}\left\{\pm \chi_0\vec\chi + \frac{1}{2}
[\vec\chi\times\vec\chi]\right\}, ~~ 
\bar{\bf J}_{2,4} = \frac{\ri}{2}\left\{\pm \bar\chi_0\vec{\bar\chi} + \frac{1}{2}
[\vec{\bar\chi}\times\vec{\bar\chi}]\right\}\label{curr2}
\eea
 The fields $\chi, \bar\chi$ stand for real (Majorana) fermions.  

 Eq.(\ref{Ising}) describes a critical Ising model; the corresponding  excitations 
 are  gapless and non-magnetic; they  appear in the  sectors with both parities. 

  Let me say several words about the O(3) Gross-Neveau model (\ref{O3}). Though  Majorana fermion description presents some advantages, the staggered magnetization components as well as their product (\ref{V}) are nonlocal with respect to these fermions. It turns out that the latter  interaction can be expressed in terms of order and disorder parameter operators of the eight Ising models corresponding to each Majorana fermion species. Since we are  interested in the case when $|\delta J| \leq M$, we have to recast the perturbation in terms of the nonchiral fields of models $H_{\pm}$. This can be done using Abelian bosonization representation for individual chains and the correspondence between $C=1$ theory and two critical Ising models (see \cite{Book}, \cite{shelton}, \cite{me}). The net result for the density of the perturbation Hamiltonain obtained after some algebra is  

\bea
&&{\cal H}_2 = \delta J({\bf N}_1 + {\bf N}_3)({\bf N}_2 + {\bf N}_4) = \nonumber\\
&&\delta J(\s_0^+\s_0^-)[(\s_1\s_2\s_3)^+(\s_1\s_2\s_3)^- - 3(\mu_1\mu_2\mu_3)^+(\mu_1\mu_2\mu_3)^- + ...] + \nonumber\\
&&\delta J(\mu_0^+\mu_0^-)[3(\s_1\s_2\s_3)^+(\s_1\s_2\s_3)^- + (\mu_1\mu_2\mu_3)^+(\mu_1\mu_2\mu_3)^- + ...] \label{stagint}
\eea
where the dots stand for the terms which do not have finite averages at  $\delta J$. $\s_a^{\pm}, \mu_a^{\pm}$ ($a = 0,1,2,3$ are order and disorder parameter operators  of the Ising models associated with the Majorana fermions $\chi_a,\bar\chi_a$ from $+$ and $-$ sectors. As we shall demonstrate later, this interaction leads to confinement of spin-1/2 particles. 

The dimerization order parameter is 
\bea
&&({\bf N}_1 - {\bf N}_3)({\bf N}_2 - {\bf N}_4) = \nonumber\\
&&(\s_0^+\s_0^-)[(\s_1\s_2\s_3)^+(\s_1\s_2\s_3)^- + 3(\mu_1\mu_2\mu_3)^+(\mu_1\mu_2\mu_3)^- + ...] + \nonumber\\
&&(\mu_0^+\mu_0^-)[- 3(\s_1\s_2\s_3)^+(\s_1\s_2\s_3)^- + (\mu_1\mu_2\mu_3)^+(\mu_1\mu_2\mu_3)^- + ... ]\label{dimer}
\eea 
 As I have said, the terms in the square brackets in Eq.(\ref{stagint}) have nonzero vacuum averages even at $\delta J =0$. The unperturbed model has four ground states: with $\la (\s_1\s_2\s_3)^{\pm}\ra \neq 0$ (the $\s^{\pm}$ vacua) and $\la (\mu_1\mu_2\mu_3)^{\pm}\ra \neq 0$ (the $\mu^{\pm}$ vacua). Replacing in (\ref{stagint}) the corresponding products by their vacuum expectation values $\sim M^{3/8}$ and identifying $(\mu_0^+\mu_0^-) = \cos(\sqrt\pi\Phi_0), (\s_0^+\s_0^-) = \sin(\sqrt\pi\Phi_0)$, we get 
\bea
&&{\cal H}_2 \sim (\delta J) M^{3/4}\cos[\sqrt\pi\Phi_0 + \tan^{-1}3] ~~ (\la \mu^{\pm}\ra \neq 0)\nonumber\\
&&\sim (\delta J) M^{3/4} \cos[\sqrt\pi\Phi_0 - \tan^{-1}1/3] ~~ (\la \s^{\pm}\ra \neq 0) \label{forms}
\eea
and 0 if for one parity  $\la\s\ra \neq 0$ and $\la\mu\ra \neq 0$ for  the other. Thus the ground state degeneracy is now reduced to two; the order (disorder) parameters in  both sectors now condense simultaneously. The $\s \rightarrow \mu$ degeneracy is not lifted, since  two forms of the potential (\ref{forms}) are equivalent under the uniform shift 
$\sqrt\pi\Phi_0 \rightarrow \sqrt\pi\Phi_0 + \pi/2$ which does not affect the gradient term $(\p_{\mu}\Phi_0)^2$. 
 Perturbation (\ref{V}) couples the critical Ising model sectors with different parity; being projected on the state with $\la\mu^{+}\ra, \la\mu^-\ra \neq 0$  the resulting Hamiltonian  becomes  the sine-Gordon one:
\bea
&&{\cal H}_{SG} = \frac{\ri}{2}\bar\chi^+_0\tau^3\p_x\chi^+_0 + \frac{\ri}{2}\bar\chi^-_0\tau^3\p_x\chi^-_0 + AM^{3/4}(\delta J)(\s^+_0\s^-_0 + 3\mu^+_0\mu^-_0) = \nonumber\\
&& \frac{1}{2}[\Pi_0^2 + (\p_{x}\Phi_0)^2] + \tilde A (\delta J) M^{3/4}
\cos(\sqrt\pi\Phi_0 + \tan^{-1}3)
\eea
where $\tilde A \sim 1$ is a numerical coefficient, $\tau^3$ is the Pauli matrix  and $\Pi_0$ is the momentum density operator. The cosine term gives rise to the mass gap 
\be
\Delta \sim (\delta J)^{4/7}M^{3/7} \label{massd}
\ee
 The average  dimerization is 
\bea
&&\epsilon \equiv \la({\bf N}_1 - {\bf N}_3)({\bf N}_2 - {\bf N}_4)\ra = \nonumber\\
&& \sim M^{3/4}\la\cos[\sqrt\pi\Phi_0 - \tan^{-1}3]\ra ~~ (\la \mu \ra \neq 0) \nonumber\\
&& \sim - M^{3/4}\la\cos[\sqrt\pi\Phi_0 + \tan^{-1}1/3]\ra ~~ (\la \s \ra \neq 0) 
\eea 
This average   does not depend on the choice of vacuum (that is whether $\s$ or $\mu$ are in the condensate):
\bea
\epsilon \sim (\delta J)^{1/7}M^{6/7}  
\eea
Thus the dimerization changes its sign when $\delta J$ goes through zero  following a power law dependence. The exponent is quite small and it is possible that in the limit of infinite number of chains the transition is the 1st order, as in the isotropic CF model \cite{batista}. I believe that this is indicative of the physics behind the deconfinement: it occurs on the boundary between two Valence Bond crystalline orders.

\begin{figure}[ht]
\begin{center}
\epsfxsize=0.45\textwidth
\epsfbox{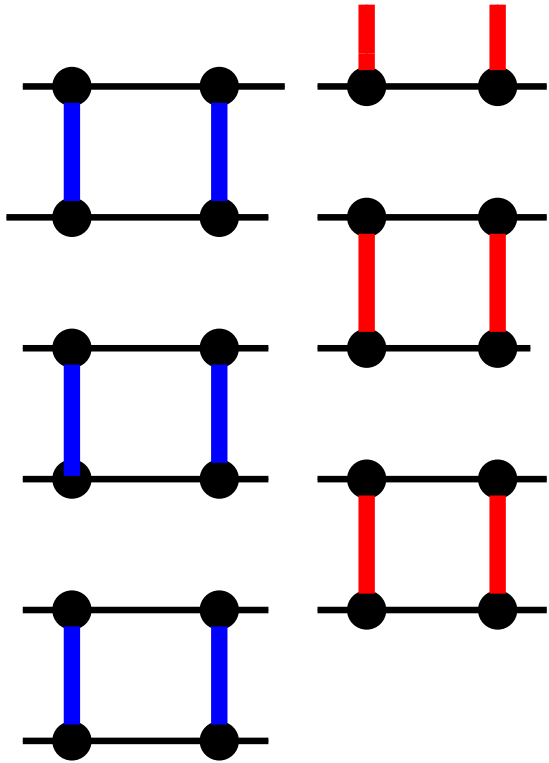}
\end{center}
\caption{ The dimerization patterns for two VBC orders. The dots are spins; the strong bonds are depicted in color.  }
\end{figure}

\begin{figure}[ht]
\begin{center}
\epsfxsize=0.65\textwidth
\epsfbox{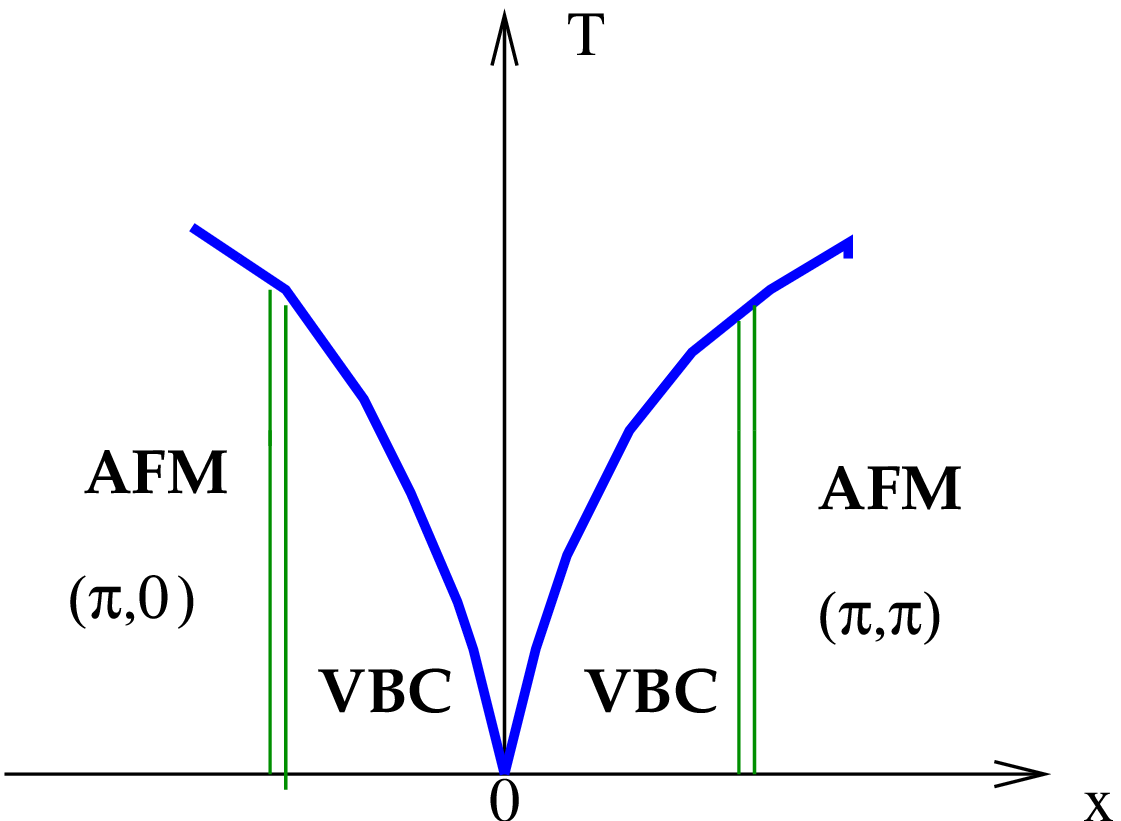}
\end{center}
\caption{ The schematic phase diagram for the CF model. The bold green lines denote the 1st order phase transitions between the antiferromagnetic and the dimerized states. $x = \delta J/M$. The dimerization changes sign at $x =0$. }
\end{figure}

\section{Confinement of Spinons}

 The problem of kinks confinement in the potential (\ref{stagint}) is somewhat peculiar differing from the standard confinement problems studied by various authors \cite{mussardo}, \cite{affleck}, \cite{fonseca}. In our case the string tension is provided by the $\cos[\sqrt\pi\Phi_0], \sin[\sqrt\pi\Phi_0]$ terms and becomes energy dependent. Though it is not directly related to the main topic of the paper, it is interesting enough on its own right. So  I will spent some time discussing this problem.
 
Let us consider a kink interpolating between the $\s$- and $\mu$-vacua in the $+$ sector ($[\mu\s]^{+}$ kink). It carries an isotopic (spin) index $\alpha = \pm 1/2$. According to (\ref{stagint}), the creation of such a kink  will  lead to the rise of the total energy proportional to the system size. To prevent this, one has to create a similar kink in the $-$ sector. If we have two kinks $\mu\s$ - one centered at $x=x_1$ and the other one at $x=x_2$, the effective potential for the $\sqrt\pi\Phi = \sqrt\pi\Phi_0 + \tan^{-1}3$ field is 
\bea
\tilde A(\delta J)M^{3/4}\left\{\theta(x_1 -x)\cos[\sqrt\pi\Phi] + \theta(x - x_2)\sin[\sqrt\pi\Phi]\right\}
\eea
The kinks are heavy particles whose mass $M$ far exceeds the masses of excitations of the $\Phi_0$ field $\sim \Delta$. This justifies the approximation which takes the kink configuarations  as step functions. At $x < x_1$ field $\Phi$ is locked at $\sqrt\pi$, at $x > x_2$ it is locked 
at $3\sqrt\pi/2$. The solution in the middle is 
\[
\sqrt\pi\Phi = \frac{\pi}{2}\frac{x - x_1}{x_2 - x_1}
\]
such that the total energy difference between the vacuum without kinks and the vacuum in the presence of two kinks is 
\bea
V = B^2\Delta^2|x_{12}| + \frac{\pi}{8|x_{12}|} \label{potential} 
\eea
where $B \sim 1$ (though its numerical value can be calculated, it is not of an interest here) and 
\be
B^2\Delta^2 = \tilde A (\delta J)M^{3/4}\la\cos[\sqrt\pi\Phi]\ra  
\ee

\begin{figure}[ht]
\begin{center}
\epsfxsize=0.5\textwidth
\epsfbox{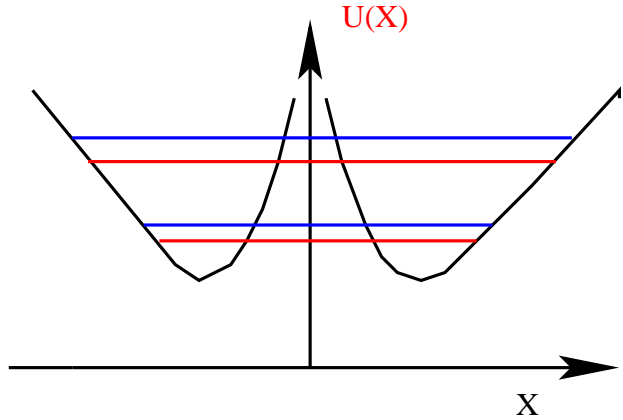}
\end{center}
\caption{ The schematic form of the confining potential (\ref{potential}). The energy levels corresponding to odd and even parity states are colored in blue and red respectively.}
\end{figure}

This energy difference provides an effective confining double-well potential in which  the $[\mu\s]^{+}-[\mu\s]^{-}$ bound states are formed. Recall that the potential does not depend on the spin configuration (this will be no longer the case when $\Delta \sim M$). Therefore the bound states form 4-fold degenerate isotopic multiplets consisting of SU(2) triplets and singlets.  The mass spectrum of kink bound states in such double-well  potential is somewhat different from the purely linear confining potential  studied in \cite{mussardo},\cite{affleck},\cite{fonseca}:
\bea
&& M_n - 2M \approx \nonumber\\
&& B\Delta\left\{\sqrt{\pi/2}  + \left[\sqrt{8/\pi}\frac{B\Delta}{M}\right]^{1/2}(n + 1/2)[1  \pm \delta(n)] + ...\right\}, ~~ n << (M/\Delta)^{2/5} \label{spec}
\eea
where 
\[
\delta(n) \sim \exp\left[- (\pi/2)^{17/8}(M/B\Delta)^{3/4}\frac{1}{\sqrt{n + 1/2}}\right]
\]
and 
\bea
M_n - 2M \approx B\Delta(n^2B\Delta/M)^{1/3}, ~~ n >> (M/\Delta)^{2/5}
 \label{spec1}
\eea
Not all these bound states are stable; the particles with masses greater than $2M + \Delta$ can decay into particles with smaller masses emitting excitations of the $\Phi_0$ field. Therefore particles with $n > \sqrt{M/\Delta}$ are unstable. Since the power 1/2 is quite close to 2/5, the mass sequence of  Eq.(\ref{spec1}) is never reached. 
 
 As far as kink-antikink states are concerned, for them the string potential is not repulsive, but attractive and they are expelled from the spectrum. 

\section{Conclusion}

 The study of the four chain case gives reasons  to believe that the infinite system has a  phase diagram presented on Fig. 4. The deconfinement point appears in the middle of the VBC phase, as in the dimer models considered in \cite{fradkin},\cite{sent}. The excitation spectra are different, however. The spinless modes of the dimer models have a quadratic spectrum $\omega \sim k^2$; such spectrum cannot emerge in CF model due to the (1+1)-dimensional Lorentz invariance. Thus I conclude that the D-point of CF model belongs to a universality class different from the universality class of the Roksar-Kivelson critical point. 

 In conclusion to this paper I  would like to point out a rather curious parallel between the problem of frustrated magnetism and another long standing problem of condensed matter physics, namely the problem of heavy fermion state formation in rare earth compounds. In these compounds 
magnetic moments of highly localized electrons belonging to rare earth ions $f$ shells interact with delocalized electrons from the broad conduction band. The  problem is discussed in the literature in terms of competition between the Kondo screening and the induced interspin interaction (the so-called RKKY interaction). On the formal level this competiton looks exactly like the competition between the less relevant current-current interaction in model (\ref{model}) and the more relevant interaction of staggered magnetisations (\ref{V}). Indeed, the energy scale generated by the Kondo screening (Kondo temperature) is exponentially small in the coupling constant as in (\ref{Mass}) and the energy scale generated by the RKKY interaction is proportional to the square of the spin-fermion 
coupling. Though one would expect that $\exp(-1/g)$ is always much smaller than $g^2$, there is a vast class of materials where the Kondo screening is manifest at temperatures much larger than the temperature of magnetic ordering. One possible explanation is that the RKKY interaction with its oscillatory behavior, is highly frustrated which leads to cancellations similar to the one considered in this paper. 


 I am grateful to O. Starykh, I. Zaliznyak, A. Nersesyan and F. H. L. Essler for discussions and  interest to the work. I  acknowledge the support from 
US DOE under contract number DE-AC02 -98 CH 10886.

\end{document}